# Voltage Controlled Memristor Threshold Logic Gates

Akshay Kumar Maan
Queensland Microelectronic Facility
Griffith University
Queensland 4111, Australia
Email: akshay.maan@gmail.com

Alex Pappachen James
School of Engineering, Nazabayev University
Astana, Kazakhastan
Web: www.biomicrosystems.info/alex
Email: apj@ieee.org

*Abstract*—In this paper, we present a resistive switching memristor cell for implementing universal logic gates. The cell has a weighted control input whose resistance is set based on a control signal that generalizes the operational regime from NAND to NOR functionality. We further show how threshold logic in the voltage-controlled resistive cell can be used to implement a XOR logic. Building on the same principle we implement a half adder and a 4-bit CLA (Carry Look-ahead Adder) and show that in comparison with CMOS-only logic, the proposed system shows significant improvements in terms of device area, power dissipation and leakage power.

## I. Introduction

The discovery of the physical memristor, the fourth basic circuit element, as theorised by Leon Chua in 1971 [1], by Hewlett-Packard Laboratories [2], has refocussed the research on memristive systems. The memristor device exhibits a resistive switching phenomenon in which a dielectric suddenly changes its (two terminal) resistance under the action of a strong electric field or current. Since the change in resistance is non-volatile and reversible, applications containing such devices lay within memory and computing applications [3]–[6].

The circuits that are inspired from firing of neurons and weighted inputs referred to as threshold logic can be implemented with memristors due to their varied resistive states that are useful as weights. There are, in general, several challenges to threshold logic implementation in silicon, including a limited number of inputs per gate, limited generalization of the cell, and difficulties with integration with conventional logic families [7]. Recent developments in memristor technology [1] have stimulated renewed interest in using switching devices for developing threshold logic gates. However, although these have smaller chip areas and lower leakage power than conventional CMOS logic gates, they often have higher power dissipation for threshold logic gates.

The key element of the proposed CMOS inverter/memristor circuits is a *memristive* [1], [8], [9] device that exhibits resistance-switching [10], [11], [12], [13]. We propose a memristor-based resistive threshold logic gate family that progresses from previous work [3] on memristive gates with respect to the simplified cell structure with the ability to control the gate operations through external input control signals. This strategy is inspired from the cognitive ability of human brain which results from the brain's ability to program and reuse similar neural structures through formation of appropriate neural networks. In the proposed design, the resistive switching devices [14], are used as programmable weights to the inputs, while the CMOS inverter behaves as a threshold logic device. The weights are programmed via the resistive switching phenomenon of the memristor device. We show that resistive switching makes it possible to use the same cell architecture to work in the NAND, NOR or XOR configuration, and can be implemented in a programmable array architecture. We hypothesise that if such circuits are developed in silicon that can be programmed and reused to generate different logic gate functionalities, we will be able to move a step closer towards the development of low power and large scale threshold logic applications.

### A. Proposed Logic

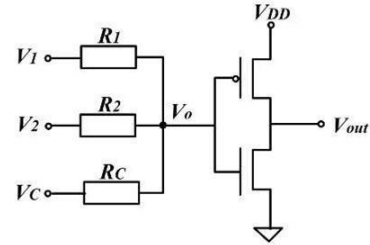

Fig. 1. Resistive memory threshold logic cell circuit consisting of resistance switching devices $R_1$, $R_2$ and $R_c$, and a CMOS inverter

In the proposed logic, we first design a threshold logic based cell that will act as a NOR or NAND gate based on a preset control voltage $V_C$ (Fig. 1). This cell forms the unit which, along with the resistive switching property of the memristors ($R_1$, $R_2$, $R_C$) involved, can be used to design an entire logic family of gates. We call it resistive switching threshold logic and use it to design an XOR gate and larger XOR circuits: a full adder and a 4-bit Carry Look-ahead Adder (LCA). Figure 1 shows the design of the proposed memory threshold logic cell that utilizes two input memristors $R_1$ and $R_2$, connected to the inputs $V_1$ and $V_2$, respectively. In general, $V_i$ is the input voltage corresponding to $i_{th}$ input terminal, where $i$ can have values $1, 2, 3, \ldots, N$. The control voltage $V_c$ connected to resistor $R_c$ is used to program the gate to NAND, NOR or XOR functionality. In case of a 2-terminal cell in-order to get the NOR functionality the control voltage $V_c$ should be greater than 0.5 V (logic 1) and for NAND gate it should be below 0.5 V (logic 0), if the input voltage values are 0V and 1V for low and high voltage levels. In order to get the XOR functionality, $V_c$ should be the NOR output of inputs $V_1$ and $V_2$. For NAND and NOR the input resistance is set as

$R_1 = R_2 = R_c = R$, while for XOR configuration $R_c$ changes according to the control voltage $V_c$. The CMOS inverter, with an inverter threshold $V_{th}$, is used to convert the $V_o$ voltage to the binary state $V_{out}$, reflective of the gate operation. All voltages share a common ground.

### B. Threshold Logic

A linear threshold logic gate (LTG) has the following transfer function:

$$f(x_1, \ldots, x_n) = \begin{cases} 1 & \text{if } \sum_{i=1}^{n} w_i x_i >= T \\ 0 & \text{otherwise} \end{cases} \quad (1)$$

where $x_i$ is a Boolean input variable, $w_i$ is an integer weight of the corresponding input $i$, and the threshold $T$ is an integer number. In the proposed design we implement the weights $w_i$ using the resistive switching property of the memristor device, $x_i$ are the inputs available to the gates and the threshold $T$ is implemented using a CMOS inverter [8].

### C. Memristor

On applying an initial threshold voltage memristive devices are brought into ON state upon which they show linear conductance with a configurable slope $1/R$ [8]. Thus they are especially suitable for implementing weights in the proposed threshold logic circuits. The memristor model used for simulation is the spice model that was proposed by [15]. This device has a large $R_{OFF}/R_{ON}$ ratio ($10^6$) while still retaining a relatively low switching time (about $10ns$).

From Figure 2 we can see that when voltages $3.5V$ and $-3.5V$ are applied across the positive and negative terminals of the memristor, we get a high resistance state (which we denote with $R_H$) and if the reverse voltages ($-3.5V$ and $3.5V$) are applied across its positive and negative terminals, we get low resistance state (which we denote with $R_L$). This voltage levels, $+/-3.5V$, are used to set high and low resistance states of memristor and are represented as $+/-V_{TR}$, the training voltage [7]. If we use any voltage in between the training voltages the resistance state of the memristor will not change. Hence we can use these voltage levels for the working of the cell and will refer to them as the testing voltage $V_{TE}$ which can be $1V$ for logic high and $0V$ for logic low.

### D. NAND/NOR Threshold Logic

From Figure 1, the output voltage of the $N$ input resistive divider circuit is given by:

$$V_o = \frac{V_c/R_c + \sum V_i/R_i}{1/R_i + 1/R_c} \quad (2)$$

For NAND, NOR and XOR operation we keep the resistances of the inputs to a high resistance $R_H$ state. For NAND/NOR operation, $R_c$ also remains at high resistance. To configure the circuit as NAND gate the control voltage $V_c$ is logic high, whereas for NOR functionality it is logic low.

Since for NAND and NOR configuration all resistors are kept equal, for a 2 input cell Eq. 2 changes to: $V_o = (V_1 + V_2 + V_c)/3$. While the threshold operation using the inverter can be represented as:

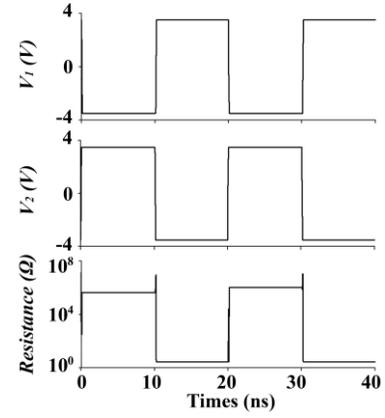

Fig. 2. Spice simulation result shows the switching of resistance state of the memristor when voltages $V_1$ & $V_2$ applied across it

TABLE I. TRUTH TABLE FOR NAND AND NOR OPERATION BASED CONTROL VOLTAGE $V_c$

| $V_1$ | $V_2$ | $V_c$ NAND | $V_c$ NOR | $R_c$ $V_c=0$ | $R_c$ $V_c=1$ | $V_o$ $V_c=0$ | $V_o$ $V_c=1$ | $V_{out}$ NAND | $V_{out}$ NOR |
|---|---|---|---|---|---|---|---|---|---|
| 0 | 0 | 0 | 1 | $R_H$ | $R_H$ | 0 | < 0.5 | 1 | 1 |
| 0 | 1 | 0 | 1 | $R_H$ | $R_H$ | < 0.5 | > 0.5 | 1 | 0 |
| 1 | 0 | 0 | 1 | $R_H$ | $R_H$ | < 0.5 | > 0.5 | 1 | 0 |
| 1 | 1 | 0 | 1 | $R_H$ | $R_H$ | > 0.5 | > 0.5 | 0 | 0 |

$$V_{out} = \begin{cases} 1 & \text{if } V_o < V_{th} \\ 0 & \text{otherwise} \end{cases} \quad (3)$$

As shown in Table I NAND logic is achieved when the control voltage $V_c$ is logic 0 and NOR logic is obtained when the control voltage $V_c$ is logic 1.

### E. XOR Threshold Logic

Figure 3 shows the XOR configuration of resistive memory threshold logic. In order to get the XOR configuration resistors $R_1$ and $R_2$ should be at high resistive state $R_H$ and $R_c$ should change its resistance according to the control voltage $V_c$ based on Eq. 4.

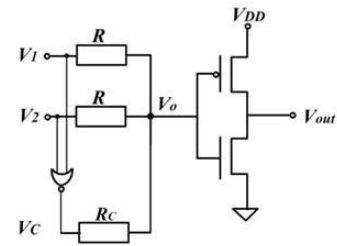

Fig. 3. XOR configuration of resistive memory threshold logic.

TABLE II. TRUTH TABLE FOR XOR OPERATION BASED CONTROL VOLTAGE $V_c$ AND CORRESPONDING CHANGE IN $R_c$. HERE $V_c$ IS THE NORED OUTPUT OF $V_1$ AND $V_2$

| $V_1$ | $V_2$ | $V_c$ | $R_c$ | $V_o$ | $V_{out}$ |
|---|---|---|---|---|---|
| 0 | 0 | 1 | $R_L$ | > 0.5 | 0 |
| 0 | 1 | 0 | $R_H$ | < 0.5 | 1 |
| 1 | 0 | 0 | $R_H$ | < 0.5 | 1 |
| 1 | 1 | 0 | $R_H$ | > 0.5 | 0 |

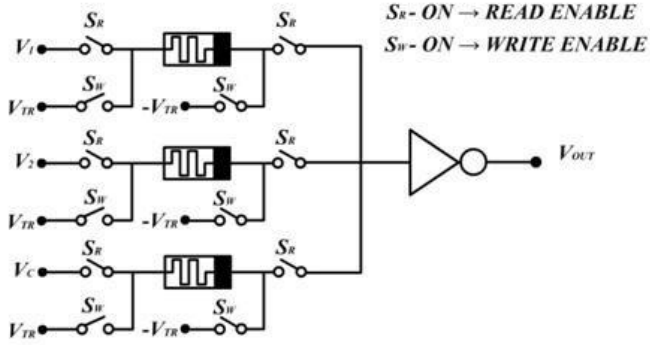

Fig. 4. Basic hardware realisation of the cell incorporated with training circuits.

$$R_c = \begin{cases} R_L & \text{if } V_c = 1 \\ R_H & \text{if } V_c = 0 \end{cases} \quad (4)$$

Here the control voltage itself is the NOR output of the inputs, so that we get the truth table as shown in Table II.

In order to realize the proposed NAND, NOR and XOR gates as a programmable circuit, we need the basic cell of Figure 1 in circuit form. Figure 4 shows the basic cell incorporated with programming circuits. Data is stored on a memristor (and in memories in general) during the write phase (cycle) and read during the read phase. The switches $S_W$ in Fig. 4 will close during writing phase and the memristors will be brought to their required resistance states. Switches $S_R$ will close during reading phase when required inputs (such as $V_1$ and $V_2$ generated through testing voltage $V_{TE}$) are applied to the cell.

## II. RESULTS AND SIMULATIONS

All simulations were performed using LTSpice [16]. CMOS technologies used for simulations have feature size of $0.25\mu m$ TSMC process BSIM models. The memresistive device model [15] used here has a large ROFF/RON ratio ($10^6$). The power supply ($V_{DD}$) is kept at $1V$ and the logic input ($V_i$) levels are $0V$ for logic low, and $1V$ for logic high. The threshold voltage of the inverter $V_{th}$ is kept at $0.5V$.

### A. NAND/NOR

Figure 5 shows the timing diagram for NAND and NOR operation when input pulse $V_1$ of $1\mu s$, 50% duty cycle and $V_2$

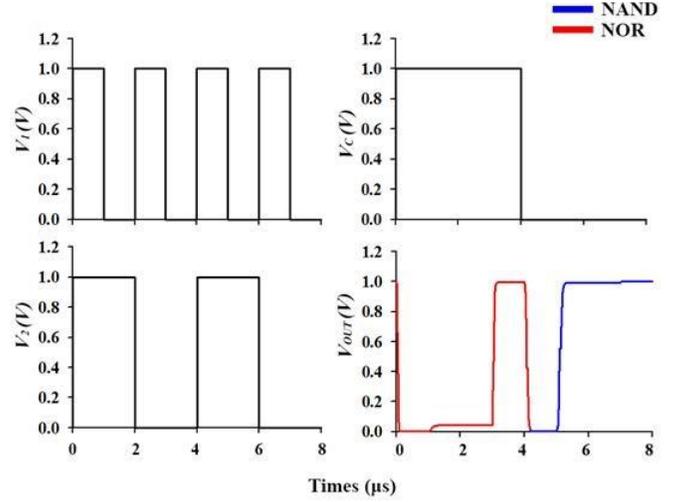

Fig. 5. Timing diagram showing the NAND/NOR operation of the proposed memristor threshold logic.

TABLE III. PERFORMANCE COMPARISON OF PROPOSED XOR LOGIC WITH CMOS AT SAME OPERATIONAL SPEED OF 1 MHZ

| Logic Family | Area ($\mu m^2$) | Power Dissipation ($\mu W$) |
|---|---|---|
| CMOS | 19.4 | 0.42 |
| RMTL | 9.4 | 0.18 |
| RMTL including training circuits | 100.50 | 33.43 |

of $2\mu s$, 50% duty cycle and $V_c$ of $4\mu s$, 50% duty cycle are applied to the circuit to get the NAND and NOR functionalities. This response is from a 2 input single cell.

### B. XOR

Figure 6 shows the behaviour of a 2 input XOR cell when input pulses $V_1$ and $V_2$ of $1\mu s$ and $2\mu s$, resp., with 50% duty cycle are applied as testing voltages. Here the control voltage $V_c$ is the NORed output of the testing voltages. We used control signals generated separately by a control circuit which were then used to program the cell as per the input testing voltages $V_1$ and $V_2$.

Table III shows the performance comparison of the proposed XOR circuit with a conventional CMOS implementation. This study is based on the simulations done on a 2 input XOR cell. All the reported values are based on the cell devices parameters only, the control circuitry is not considered for this study. We note that the area, power dissipation and leakage power all show much improvement in performance over the conventional CMOS circuits.

Figure 7 shows an implementation of a half adder circuit using the proposed method. The control voltage which is the NOR output of input voltages is implemented as described above. The performance comparison of proposed RMTL gates with CMOS implementation in applications like the half adder and CLA (Carry Look-ahead Adder) is shown in Table IV. Here the gains of RMTL over CMOS circuits in both area and power dissipation for both half adder and 4-bit CLS is due to RMTLs logic implementation using memristors.

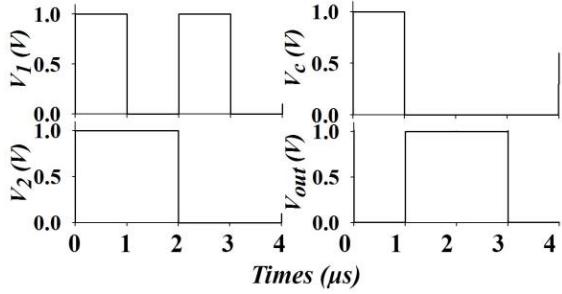

Fig. 6. Simulated result of the XOR cell when the testing voltages $V_1$ and $V_2$ and the control voltage $V_c$ are applied.

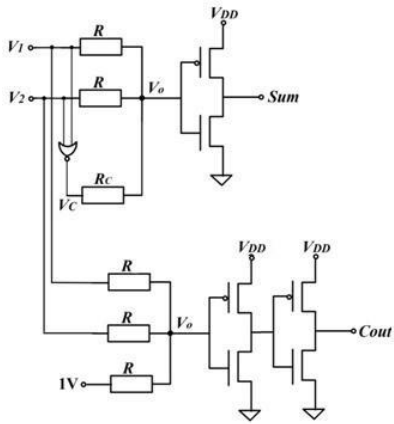

Fig. 7. Full adder using resistive memory threshold logics.

### III. CONCLUSIONS

The proposed threshold logic programmable gate arrays represent a practical application of quantized resistive memory devices in the design of generalized logic gates. This logic family of gates is inspired by the cognitive logic circuits of the human brain, and is an example of mimicking neuronal logic circuits. The philosophical argument for the need for a cognitive logic family is presented, and in comparison with conventional CMOS logic, our devices benefit from smaller area and lower power dissipation. In addition, since the switching devices are silicon based, the integration of the proposed logic with CMOS logic gates is practically feasible, and can be used in combination to improve the performance of existing digital logic designs.

TABLE IV. PERFORMANCE COMPARISON OF A 1-BIT HALF ADDER AND A 4-BIT CLA USING CMOS AND RMTL

| Application | Logic Family | Area ($\mu m^2$) | Power Dissipation ($\mu W$) |
|---|---|---|---|
| 1Bit Full Adder | CMOS | 32.92 | 0.82 |
|  | RMTL | 18.80 | 0.58 |
| 4Bit CLA | CMOS | 1261 | 99.64 |
|  | RMTL | 244 | 9.612 |

All comparisons are made without considering the control circuits